\documentclass[preprint,showpacs,preprintnumbers]{revtex4}
\usepackage{graphicx}
\usepackage{dcolumn}
\usepackage{bm}
\usepackage{amsmath}
\usepackage{amsfonts}
\usepackage{amssymb}

\input{tcilatex}

\begin{document}

\title{Scanning probe microscopy of thermally excited mechanical modes of an
optical microcavity}
\author{T. J. Kippenberg\thanks{%
Present address: Max Planck Institute of Quantum Optics, 85748 Garching,
Germany.}, H. Rokhsari, K.J. Vahala}
\email{vahala@its.caltech.edu}
\affiliation{California Institute of Technology, Department of Applied Physics, Pasadena,
CA 91125, USA. }

\begin{abstract}
The resonant buildup of light within optical microcavities elevates the
radiation pressure which mediates coupling of optical modes to the
mechanical modes of a microcavity. Above a certain threshold pump power,
regenerative mechanical oscillation occurs causing oscillation of certain
mechanical eigenmodes. Here, we present a methodology to spatially image the
micro-mechanical resonances of a toroid microcavity using a scanning probe
technique. The method relies on recording the induced frequency shift of the
mechanical eigenmode when in contact with a scanning probe tip. The method
is passive in nature and achieves a sensitivity sufficient to spatially
resolve the vibrational mode pattern associated with the thermally agitated
displacement at room temperature. The recorded mechanical mode patterns are
in good qualitative agreement with the theoretical strain fields as obtained
by finite element simulations.
\end{abstract}

\maketitle

%\email{}

The work of V.B. Braginsky\cite{Braginsky2001} predicted that due to
radiation pressure the mechanical mirror-eigenmodes of a Fabry-P\'{e}rot
(FP) resonator can couple to the optical modes, leading to a parametric
oscillation instability. This phenomenon is characterized by regenerative
mechanical oscillation of the mechanical cavity eigen-modes. Significant
theoretical studies have been devoted to this effect in the context of the
laser gravitational wave observatory (LIGO) (\cite{Braginsky2001, Kells2002}%
), as it potentially impedes gravitational wave detection. Whereas in
macroscopic resonators the influence of radiation pressure is weak and only
appreciable at high power levels\cite{Dorsel1983}, the mutual coupling of
optical and mechanical modes is significantly enhanced in optical
microcavities (such as silica microspheres\cite{Braginsky1989}, microdisks
or toroids\cite{Armani2003}) which simultaneously exhibit ultra-high-Q
optical modes \emph{and} high-Q mechanical modes in the radio-frequency
range. The combination of high optical power and small mechanical mass and
dissipation leads to threshold levels in the micro-Watt regime for
regenerative acoustic oscillations (i.e. parametric oscillation
instability), making it the dominant micro-cavity nonlinear optical effect
as reported previously in toroid microcavities\cite{Kippenberg2005,
Rokhsari2005, Carmon2005}.

In this letter, we report a novel scanning-probe technique, which allows
direct \emph{spatial imaging} of the amplitude of the micro-mechanical modes
of a microcavity associated with their thermally driven displacement at room
temperature. The method is based on the induced resonance shift by a
scanning probe tip, whose influence on the mechanical oscillator's resonance
frequency is detected optically via the light transmitted past the
microcavity. This technique is passive in nature, and reaches a sensitivity
which is sufficient to detect the minute amplitude of the thermally driven
mechanical modes. Initial demonstrations of this method show very good
agreement between the mechanical mode distribution obtained by
scanning-probe spectroscopy and finite-element modeling. Besides providing
insight into the spatial pattern of the mechanical modes of an optical
microcavity, this technique should provide a useful tool for the study of
other micromechanical or nano-mechanical resonators\cite{Craighead2000}.

The experimental scenario is depicted schematically in Figure 1. It consists
of a standard geometry in which a pump wave is coupled from a waveguide (a
tapered optical fiber\cite{Cai2000}) to an ultra-high-Q microcavity mode of
a toroid microcavity on a chip \cite{Armani2003}. In addition to their
excellent optical properties, this microcavity geometry - owing to its free
hanging silica membrane supporting the toroidal periphery - possesses high-Q
micromechanical resonances. The inset in figure 3 shows the first two ($%
n=1,2 $) mechanical modes of the structure, calculated using finite element
modeling. The modes are rotationally symmetric (i.e. their azimuthal mode
number being $m=0$). As evident from the finite element simulation, the
motion of the first and second order flexural mode is primarily in the
out-of plane direction (of the toroid and disk).

On resonance the high buildup of light within the cavity leads to an
increase in radiation pressure, expanding the cavity (optical round trip)
and thereby coupling the optical mode to the mechanical eigenmodes of the
cavity as described in Refs\cite{Kippenberg2005, Rokhsari2005, Carmon2005}.
The mutual coupling of the mechanical and optical mode is described by the
following set of equations:

\begin{equation}
\frac{\mathrm{d}^{2}}{\mathrm{d}t^{2}}x-\frac{\omega _{m}}{Q_{m}}\frac{%
\mathrm{d}}{\mathrm{d}t}x+\omega _{m}^{2}x=K_{om}\frac{{|a|}^{2}}{T}
\end{equation}%
\begin{equation}
\frac{\mathrm{d}}{\mathrm{d}t}a=-\frac{1}{\tau }a+\mathrm{i}\left( \Delta
\omega +K_{mo}x\right) a+\mathrm{i}s\sqrt{\frac{1}{\tau _{ex}}}  \label{eq2}
\end{equation}

The first equation describes the mechanical eigenmode, where $\omega _{m}$
is the mechanical frequency, ${|x_{m}|}^{2}$ is normalized to mechanical
energy i.e. ${|x_{m}|}^{2}=\sum_{i=r,z,\Theta }\int \epsilon _{i}\sigma _{i}%
\mathrm{d}V\equiv m_{eff}\cdot \omega _{m}^{2}\cdot r^{2}$, which decays
with the lifetime $\frac{1}{\tau _{m}}$ (i.e. $Q_{m}=\omega _{m}\tau _{m}$).
Correspondingly $|a|^{2}$ is the energy in the optical whispering gallery
mode ($1/T\cdot |a|^{2}$ is the power, where $T=\frac{2\pi Rn_{e\!f\!f}}{c}$
is the cavity round-trip time), which is excited with a pump laser detuned
by the amount $\Delta \omega $ from line-center. The expressions $K_{om}$
and $K_{mo}$ describe the mutual coupling of optical and mechanical
eigen-modes, and depend on the respective modes. The coupling can be mapped
to a Fabry-P\'{e}rot cavity, by considering the modes as a harmonic
oscillator with in plane (radial amplitude $r$, which modulates the cavity
pathlength) and out-of plane motion (amplitude $z$). Solving equations
(1)-(2) shows that the radiation pressure causes the mechanical oscillator
to experience: (1) a change in rigidity; (2) the addition of a velocity
dependent term (providing either a viscous force or an accelerating force),
i.e. 
\begin{equation}
\frac{\mathrm{d}^{2}}{\mathrm{d}t^{2}}x-\left( \frac{\omega _{m}}{Q_{m}}%
+\Delta \beta _{L}\right) \frac{\mathrm{d}}{\mathrm{d}t}x+\left( \omega
_{m}^{2}+\frac{\Delta k_{L}}{m}\right) x=0
\end{equation}%
The approximate solutions are (for $\omega _{m}\ll \omega _{0}/Q$): $\Delta
\beta _{L}=\tau \frac{\mathrm{d}F}{\mathrm{d}x}$and $\Delta k_{L}=\frac{%
\mathrm{d}F}{\mathrm{d}x}.$Consequently, the laser being detuned to the blue
with respect to the cavity resonance ($\Delta \beta _{L}>0$) leads to
mechanical gain. If the mechanical gain exceeds the mechanical loss, a
parametric oscillation instability can be observed in which regenerative
mechanical oscillations occur. This phenomenon has been observed for the
first time recently in toroid microcavities and has been extensively
characterized\cite{Kippenberg2005, Rokhsari2005,Carmon2005}.

Here we investigate the interaction of a local probe (whose dimensions are
small compared to the optical microcavity) with the mechanical modes, and
demonstrate a novel method which can spatially resolve the mechanical mode
pattern associated with the thermally agitated displacement of the toroid
microcavities. In order to spatially image the mechanical modes a scanning
probe is introduced as shown in Figure 1 which contacts the free hanging
disk connected to the toroidal cavity and supported by the silicon pillar.
The scanning probe setup consisted of a silica tapered fiber tip controlled
by a piezoelectric stage (with 80 nm resolution). The tip of the scanning
probe was fabricated by $\mathrm{CO}_{2}$ laser melting and stretching of a
single mode fiber, and had a tip-diameter ca. 3 $\mu m$. The probe was
lowered onto the silica support disk (i.e. the interior region of the
toroidal microcavity), while the taper was simultaneously coupled to the
toroid. Figure 1 a,c shows an optical micrograph of the taper-cavity system
in contact with the probe (top view, side view respectively). The presence
of the probe couples the mechanical microcavity mode to the acoustical modes
of the probe. This coupling has two effects; (1) the mechanical quality ($%
Q_{m}$) factor of the microcavity structure is reduced; and (2) the
mechanical eigenfrequency ($\omega _{m}$) is changed, due to a change in
rigidity\cite{Craighead2000}. We note the similarity of this method to the
"AC" mode of an atomic force microscope\cite{Binnig1986,Rugar1991} , which
uses the change in mechanical frequency of an oscillating cantilever to
record the topology of the surface (induced by position dependent forces due
to the presence of near-field forces). However, in the present case, not the
resonant frequency shift of the probe is monitored, but rather the resonant
frequency and $Q$ of the micromechanical resonator itself. As the mechanical
cavity-motion modulates the optical power coupled to the cavity (i.e.,
cavity transmission) and thereby creates sidebands at $\omega =\omega \pm
\,\omega _{m}$ in the transmitted optical spectrum, the mechanical
oscillation frequency and Q-factor can be readily measured via the
transmitted pump power past the cavity.

If the optical pump power is low (compared to the threshold for mechanical
oscillations to occur) then the optical field acts purely as a probe of the
mechanical oscillation and does not modify the mechanical properties of the
structure (i.e. the light will not excite mechanical resonances, since $P\ll
P_{thresh}$ equivalent to $\Delta \beta _{L}$ $\ll \frac{\omega _{m}}{Q_{m}}$%
). The threshold for mechanical oscillation can be increased rapidly (and
the regime $P\ll P_{thresh}$ achieved), when biasing the coupling junction
into the overcoupled regime\cite{Spillane2003}, owing to the fact that
threshold scales inverse cubically with the optical $Q$ factor (in cases
where the mechanical frequency is smaller than the cavity bandwidth\cite%
{Kippenberg2005, Rokhsari2005}). Therefore the system can be described by a
simplified set of equations by introducing $\omega _{m}^{\ast }$ and $\Delta
\beta ^{\ast }$ (which contain the radiation pressure induced shift in
resonance and change in rigidity):

\begin{equation}
\frac{\mathrm{d}}{\mathrm{d}t} a = - \frac{1}{\tau} a + \mathrm{i} \big( %
\Delta \omega + K_{mo} \, x_T \,\mathrm{cos}\left( \omega_m t \right) \big) %
a + \mathrm{i} s \sqrt{\frac{1}{\tau_{ex}}}
\end{equation}

In this regime ($P\ll P_{thresh}$), the oscillator is only thermally driven
(i.e. the energy being equal to $k_{b}T$), causing modulation of the field
stored inside the cavity, due to the change in cavity path-length. The
solutions in this regime to the above equation are given by:

\begin{equation}
a = \sum_n \frac{s \sqrt{\tau_{ex}^{-1}} \, J_n(M) \, \mathrm{e}^{\mathrm{i}
n \omega_m t} }{- \tau^{-1} + \mathrm{i} \Delta \omega + \mathrm{i} n
\omega_m }
\end{equation}

The appearance of sidebands (at $\omega _{m}$ and harmonics) is thus
observable at the mechanical eigen-frequencies with a modulation depth $M$
which is governed by the specifics of the coupling mechanism and the
amplitude of the motion, i.e. $M=K_{mo}\cdot x_{T}$, where $x_{T}$ is the
thermal displacement noise, as given by $x_{T}=\sqrt{\frac{k_{B}T}{m\omega
_{m}^{2}}}$. The temperature $T$ in the presence of the optical probe field
is increased above the ambient temperature of 300 K by several degrees, due
to absorption of photons and subsequent heating (as evidenced from thermal
hysteresis)\cite{Rohksari2004}. Detection of light transmitted past the
microcavity exhibits modulation components at $\omega _{m}$ and harmonics,
as the transmitted light past the tapered fiber interferes with the
modulated field emerging from the cavity, i.e.

\begin{equation}
T \cong P_{in} |T_0|^2 + \sqrt{P_{in}} |T_0| \mathrm{cos}\left( \omega_m t
\right) \frac{J_0(M) \cdot 2 \,\omega_m}{{|\tau^{-1} - \mathrm{i} \Delta
\omega - \mathrm{i} \omega_m|}^2}
\end{equation}

The transmission is given by $T_{0}^{\,2}={\left\vert \frac{\tau
_{ex}^{-1}-\tau _{0}^{-1}+\mathrm{i}\Delta \omega }{\tau _{ex}^{-1}+\tau
_{0}^{-1}-\mathrm{i}\Delta \omega }\right\vert }^{2}$ and maximum modulation
depth occurs at critical coupling and for detuning of $\Delta \omega /2$,
where the slope of the Lorentzian transmission is maximum. By spectral
decomposition of the cavity transmission ($T$), the mechanical resonance
frequency and mechanical Q-factor can therefore be recorded. The inset of
Figure 2 shows the lorentzian lineshape of the first flexural mode as well
as a theoretical fit (solid line). Therefore, the transmitted pump light can
be used to optically probe both the micromechanical resonant frequency, as
well as the mechanical Q factor.

Having established a detection technique for the mechanical resonant
characteristics ($\omega _{m}$, $Q_{m}$) static probe contact measurements
were carried out. When the probe is brought in contact with the silica disk,
a reduction of the mechanical Q factor is observed, as the probe constitutes
a dissipative loss channel to the mechanical oscillator. The reduction of Q
factor increases with contact pressure of the probe. In addition, we note
that the observed decrease in $Q$ is concomitant with an increase in
mechanical eigenfrequency. Figure 2 shows $Q_{m}$ as a function of frequency 
$\omega _{m}$. As can be seen, a linear relationship between mechanical loss
($\propto Q_{m}^{-1}$) and induced frequency shift ($\Delta \omega _{m}$) is
obtained. This is non-intuitive at first sight, and not in agreement with a
damped harmonic oscillator model, which predicts that the increase in
damping (e.g. due to the tip) causes a \emph{red-shift} of the resonance
frequency, i.e. $\omega ^{\prime }=\omega \sqrt{1-\frac{4}{Q^{2}}}$.
However, the presence of the tip causes not only a change in dissipation but
also a change in the rigidity of the oscillator ($\Delta k_{P}$) which
causes a \emph{blue-shift} by $\Delta \omega _{\Delta k}=\omega _{m}\frac{%
-\Delta k}{2k}$. This effect is well known for cantilevers used for atomic
force microscopy \cite{Binnig1986,Rugar1991,CraigheadAPL}. This empirical
linear relationship of $Q_{m}$ and $\Delta \omega _{m}$ was recorded for
different modes (1,2,3) and found repeatedly for all measurements reported
in this letter.

Next \emph{spatial} probing was carried out, and the influence of the
mechanical resonant characteristics ($\omega _{m}$, $Q_{m}$) on the spatial
probe position analyzed. Figure 3 shows the results of this measurement in
which scanning is performed along the entire cross section of a toroidal
cavity with a diameter of 104 $\mu m$ (the resonant frequencies for the $n=1$
and $n=2$ modes were 6.07 MHz and 14.97 MHz). The cavity was fabricated with
a strong undercut, in order to promote the recorded amplitudes (the thermal
displacement was approximately $x^{T}\approx 1$ pm for the $n=1$ mode). As
can be observed in Fig. 3 (upper graph) for the $n=1$ mode (which in its
unperturbed state has a resonant frequency of 6.07 MHz) a continuous
decrease, followed by a plateau and then followed by an increase in the
mechanical frequencies was observed while scanning from the one outer edge
of the toroid to the opposing edge (The frequency range covered the
unperturbed value of 6.07 MHz \ to 6.17 MHz, equating to a fractional change
of $\frac{\Delta \omega _{m}}{\omega _{m}}\approx 0.016$). Similarly the
mechanical Q-factor was continually reduced, plateaud in the interior
region, followed again by an increase. The data in Fig. 3 represents the
recorded frequency shift normalized to unity. This is a first indication,
that the induced frequency shifts and Q-factor change depend critically on
the \emph{amplitude} of the mechanical oscillations, and therefore probe 
\emph{local} information about the mechanical amplitude. To confirm that
this effect is not topological in nature (e.g. due to an irregular shape of
the interior silica disk surface), and in order to establish that the
recorded shift in frequency is indeed a measure of the mechanical amplitude,
thedependencies for the $n=2$ mode were measured. The result for the $n=2$
mode is shown in Fig. 3b, and is clearly distinct from the $n=1$ mode. These
measurements were obtained repeatedly on different samples, with different
tips. As evident the mechanical frequency is perturbed maximally at the
point of maximum mechanical amplitude and decreases to zero in the interior
of the toroid, as well as at the outer edge. This clearly indicates that the
observed frequency shift is not related to the surface-topology, but in fact
provides direct information on the \emph{local} mechanical oscillation
amplitude. In order to make a qualitative comparison of the recorded
frequency shift and the actual mechanical motion, the numerically calculated
amplitude ( in the z-direction) is superimposed in figure 3. The
calculations employed finite element modeling of the actual geometry (as
inferred by SEM). \ The theoretically modeled amplitudes were scaled to
unity in the vertical direction. We note that the position of maximum
amplitude of the $n=1$ and $n=2$ mode, agrees very well with the finite
element simulation as does the overall shape of the curves.While a detailed
understanding of how the probe changes the resonant characteristics of the
oscillator is at present lacking, the data along with the modeling strongly
suggests that the recorded frequency shifts do directly relate to the strain
fields in the z-direction. Note that deviation of the recorded shift with
the numerical modeling is expected, due to the convolution of the finite
probe size (in the experiments this is approximately 3 $\mu m$) with the
mechanical motion.

In summary a novel method is presented which allows direct probing of the
vibrational mode patterns of a micromechanical oscillator. The method relies
on spatially recording the induced frequency shift by a scanning probe tip,
and exhibits sufficient sensitivity to detect the mode pattern of thermally
excited acoustic modes. The present results should also be applicable to
other types of micromechanical oscillators, and could provide a useful tool
in the field of MEMS/NEMS\cite{Craighead2000}, as well as to achieve tuning
of mechanical resonator modes\cite{Craighead2000},.

%\subsection{\textbf{Acknowledgements}}

\begin{acknowledgments}
This research was supported by the DARPA and the Caltech Lee Center for
Advanced Networking. TJK gratefully acknowledges receipt of a postdoctoral
fellowship from the IST-CPI.
\end{acknowledgments}

\bigskip 
\bibliographystyle{apsrev}
\bibliography{MicroscopyMechModes}

\begin{thebibliography}{15}
\expandafter\ifx\csname natexlab\endcsname\relax\def\natexlab#1{#1}\fi
\expandafter\ifx\csname bibnamefont\endcsname\relax
  \def\bibnamefont#1{#1}\fi
\expandafter\ifx\csname bibfnamefont\endcsname\relax
  \def\bibfnamefont#1{#1}\fi
\expandafter\ifx\csname citenamefont\endcsname\relax
  \def\citenamefont#1{#1}\fi
\expandafter\ifx\csname url\endcsname\relax
  \def\url#1{\texttt{#1}}\fi
\expandafter\ifx\csname urlprefix\endcsname\relax\def\urlprefix{URL }\fi
\providecommand{\bibinfo}[2]{#2}
\providecommand{\eprint}[2][]{\url{#2}}

\bibitem[{\citenamefont{Braginsky et~al.}(2001)\citenamefont{Braginsky,
  Strigin, and Vyatchanin}}]{Braginsky2001}
\bibinfo{author}{\bibfnamefont{V.~B.} \bibnamefont{Braginsky}},
  \bibinfo{author}{\bibfnamefont{S.~E.} \bibnamefont{Strigin}},
  \bibnamefont{and} \bibinfo{author}{\bibfnamefont{S.~P.}
  \bibnamefont{Vyatchanin}}, \bibinfo{journal}{Physics Letters A}
  \textbf{\bibinfo{volume}{287}}, \bibinfo{pages}{331} (\bibinfo{year}{2001}).

\bibitem[{\citenamefont{Kells and D'Ambrosio}(2002)}]{Kells2002}
\bibinfo{author}{\bibfnamefont{W.}~\bibnamefont{Kells}} \bibnamefont{and}
  \bibinfo{author}{\bibfnamefont{E.}~\bibnamefont{D'Ambrosio}},
  \bibinfo{journal}{Physics Letters A} \textbf{\bibinfo{volume}{229}},
  \bibinfo{pages}{326} (\bibinfo{year}{2002}).

\bibitem[{\citenamefont{{A. Dorsel et al.}}(1983)}]{Dorsel1983}
\bibinfo{author}{\bibnamefont{{A. Dorsel et al.}}}, \bibinfo{journal}{Physical
  Review Letters} \textbf{\bibinfo{volume}{51}}, \bibinfo{pages}{1550}
  (\bibinfo{year}{1983}).

\bibitem[{\citenamefont{Braginsky et~al.}(1989)\citenamefont{Braginsky,
  Gorodetsky, and Ilchenko}}]{Braginsky1989}
\bibinfo{author}{\bibfnamefont{V.}~\bibnamefont{Braginsky}},
  \bibinfo{author}{\bibfnamefont{M.}~\bibnamefont{Gorodetsky}},
  \bibnamefont{and} \bibinfo{author}{\bibfnamefont{V.}~\bibnamefont{Ilchenko}},
  \bibinfo{journal}{Physics Letters A} \textbf{\bibinfo{volume}{137}},
  \bibinfo{pages}{393} (\bibinfo{year}{1989}).

\bibitem[{\citenamefont{{Armani, D.K. and Kippenberg, T.J. and Spillane, S.M.
  and Vahala, K.J. }}(2003)}]{Armani2003}
\bibinfo{author}{\bibnamefont{{Armani, D.K. and Kippenberg, T.J. and Spillane,
  S.M. and Vahala, K.J. }}}, \bibinfo{journal}{Nature}
  \textbf{\bibinfo{volume}{421}}, \bibinfo{pages}{925} (\bibinfo{year}{2003}).

\bibitem[{\citenamefont{{Kippenberg, T.J. and Spillane, S.M. and Vahala,
  K.J.}}(2005)}]{Kippenberg2005}
\bibinfo{author}{\bibnamefont{{Kippenberg, T.J. and Spillane, S.M. and Vahala,
  K.J.}}}, \bibinfo{journal}{Physical Review Letters}
  \textbf{\bibinfo{volume}{95}}, \bibinfo{pages}{033901}
  (\bibinfo{year}{2005}).

\bibitem[{\citenamefont{{Rokhsari, H. and Kippenberg, T.J. and Carmon, T. and
  Vahala, K.J.}}(2005)}]{Rokhsari2005}
\bibinfo{author}{\bibnamefont{{Rokhsari, H. and Kippenberg, T.J. and Carmon, T.
  and Vahala, K.J.}}}, \bibinfo{journal}{Optics Express}
  \textbf{\bibinfo{volume}{13}}, \bibinfo{pages}{5293} (\bibinfo{year}{2005}).

\bibitem[{\citenamefont{Carmon et~al.}(2005)\citenamefont{Carmon, Rokhsari,
  Yang, Kippenberg, and Vahala}}]{Carmon2005}
\bibinfo{author}{\bibfnamefont{T.}~\bibnamefont{Carmon}},
  \bibinfo{author}{\bibfnamefont{H.}~\bibnamefont{Rokhsari}},
  \bibinfo{author}{\bibfnamefont{L.}~\bibnamefont{Yang}},
  \bibinfo{author}{\bibfnamefont{T.}~\bibnamefont{Kippenberg}},
  \bibnamefont{and} \bibinfo{author}{\bibfnamefont{K.}~\bibnamefont{Vahala}},
  \bibinfo{journal}{Physical Review Letters} \textbf{\bibinfo{volume}{94}},
  \bibinfo{pages}{223902} (\bibinfo{year}{2005}).

\bibitem[{\citenamefont{Zalalutdinov et~al.}(2000)\citenamefont{Zalalutdinov,
  Illic, Zehnder, Craighead, and Parpia}}]{Craighead2000}
\bibinfo{author}{\bibfnamefont{M.}~\bibnamefont{Zalalutdinov}},
  \bibinfo{author}{\bibfnamefont{B.~D.} \bibnamefont{Illic},
  \bibfnamefont{Czaplewski}},
  \bibinfo{author}{\bibfnamefont{A.}~\bibnamefont{Zehnder}},
  \bibinfo{author}{\bibfnamefont{H.}~\bibnamefont{Craighead}},
  \bibnamefont{and} \bibinfo{author}{\bibfnamefont{J.}~\bibnamefont{Parpia}},
  \bibinfo{journal}{Applied Physics Letters} \textbf{\bibinfo{volume}{77}},
  \bibinfo{pages}{3278} (\bibinfo{year}{2000}).

\bibitem[{\citenamefont{Cai et~al.}(2000)\citenamefont{Cai, Painter, and
  Vahala}}]{Cai2000}
\bibinfo{author}{\bibfnamefont{M.}~\bibnamefont{Cai}},
  \bibinfo{author}{\bibfnamefont{O.}~\bibnamefont{Painter}}, \bibnamefont{and}
  \bibinfo{author}{\bibfnamefont{K.}~\bibnamefont{Vahala}},
  \bibinfo{journal}{Physical Review Letters} \textbf{\bibinfo{volume}{85}},
  \bibinfo{pages}{74} (\bibinfo{year}{2000}).

\bibitem[{\citenamefont{Binnig et~al.}(1986)\citenamefont{Binnig, Quate, and
  Gerber}}]{Binnig1986}
\bibinfo{author}{\bibfnamefont{G.}~\bibnamefont{Binnig}},
  \bibinfo{author}{\bibfnamefont{C.}~\bibnamefont{Quate}}, \bibnamefont{and}
  \bibinfo{author}{\bibfnamefont{C.}~\bibnamefont{Gerber}},
  \bibinfo{journal}{Physical Review Letters} \textbf{\bibinfo{volume}{56}},
  \bibinfo{pages}{930} (\bibinfo{year}{1986}).

\bibitem[{\citenamefont{Albrecht et~al.}(1990)\citenamefont{Albrecht, Grütter,
  Horne, and Rugar}}]{Rugar1991}
\bibinfo{author}{\bibfnamefont{T.}~\bibnamefont{Albrecht}},
  \bibinfo{author}{\bibfnamefont{P.}~\bibnamefont{Grütter}},
  \bibinfo{author}{\bibfnamefont{D.}~\bibnamefont{Horne}}, \bibnamefont{and}
  \bibinfo{author}{\bibfnamefont{D.}~\bibnamefont{Rugar}},
  \bibinfo{journal}{Journal of Applied Physics} \textbf{\bibinfo{volume}{69}},
  \bibinfo{pages}{668} (\bibinfo{year}{1990}).

\bibitem[{\citenamefont{{Spillane, S.M. and Kippenberg, T.J. and Painter, O.J.
  and Vahala, K.J.}}(2003)}]{Spillane2003}
\bibinfo{author}{\bibnamefont{{Spillane, S.M. and Kippenberg, T.J. and Painter,
  O.J. and Vahala, K.J.}}}, \bibinfo{journal}{Physical Review Letters}
  \textbf{\bibinfo{volume}{91}}, \bibinfo{pages}{art. no.}
  (\bibinfo{year}{2003}).

\bibitem[{\citenamefont{Rokhsari et~al.}()\citenamefont{Rokhsari, Spillane, and
  Vahala}}]{Rohksari2004}
\bibinfo{author}{\bibfnamefont{H.}~\bibnamefont{Rokhsari}},
  \bibinfo{author}{\bibfnamefont{S.}~\bibnamefont{Spillane}}, \bibnamefont{and}
  \bibinfo{author}{\bibfnamefont{K.}~\bibnamefont{Vahala}},
  \bibinfo{journal}{Applied Physics Letters}  (????).

\bibitem[{\citenamefont{Zalalutdinov et~al.}(2003)\citenamefont{Zalalutdinov,
  Aubin, Pandey, Zehnder, Rand, Craighead, Parpia, and Houston}}]{CraigheadAPL}
\bibinfo{author}{\bibfnamefont{M.}~\bibnamefont{Zalalutdinov}},
  \bibinfo{author}{\bibfnamefont{K.}~\bibnamefont{Aubin}},
  \bibinfo{author}{\bibfnamefont{M.}~\bibnamefont{Pandey}},
  \bibinfo{author}{\bibfnamefont{A.}~\bibnamefont{Zehnder}},
  \bibinfo{author}{\bibfnamefont{R.}~\bibnamefont{Rand}},
  \bibinfo{author}{\bibfnamefont{H.}~\bibnamefont{Craighead}},
  \bibinfo{author}{\bibfnamefont{J.}~\bibnamefont{Parpia}}, \bibnamefont{and}
  \bibinfo{author}{\bibfnamefont{B.}~\bibnamefont{Houston}},
  \bibinfo{journal}{Applied Physics Letters} \textbf{\bibinfo{volume}{83}},
  \bibinfo{pages}{3281} (\bibinfo{year}{2003}).

\end{thebibliography}

\ 

\bigskip

\begin{figure}[tbp]
\begin{center}
\includegraphics[
width=3.729in
]{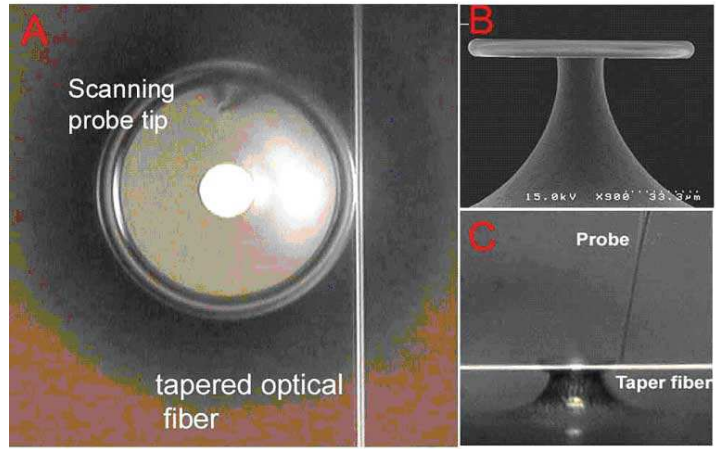}
\end{center}
\caption{ Optical micrographs of the experiment. Panel A: top view of toroid
microcavity coupled to a tapered optical fiber and contacted from the top by
a scanning probe tip. Panel B: Scanning electron micrograph (SEM) of the
toroid microcavity used to obtain the imaging results shown in Fig. 3. Panel
C: side view of a toroid cavity contacted with a scanning probe tip from the
top, and with the tapered optical fiber coupled evanescently via the side. }
\end{figure}

\begin{figure}[tbp]
\begin{center}
\includegraphics[
width=3.729in
]{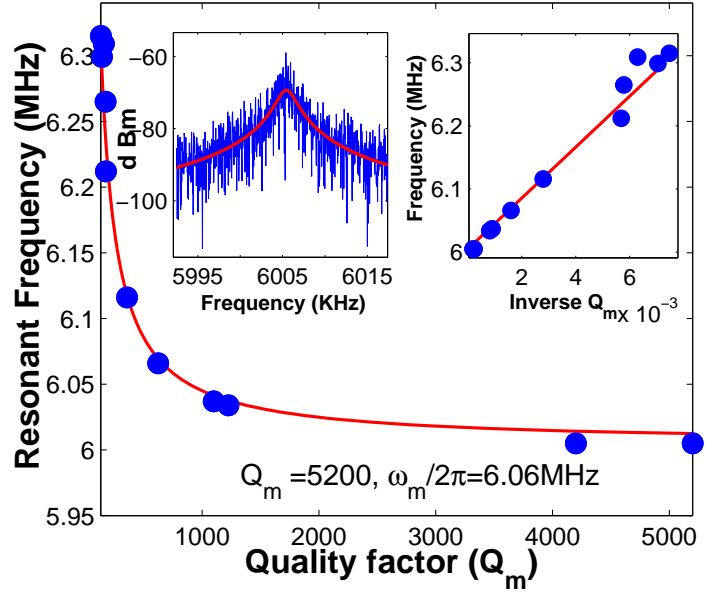}
\end{center}
\caption{Main figure: The relationship between resonant frequency of a
fundamental mechanical mode of a toroid microcavity and the mechanical $Q_{m}
$ when a probe is brought into contact with the mechanical oscillator (The
intrinsic Quality factor for this toroid was $Q_{m}=5000$ \ and it's
frequency $\protect\omega _{m}=2\protect\pi \cdot 6$ MHz).\ Right inset:\
Same data but plotted as $\protect\omega _{m}$ versus $1/Q_{m}$
demonstrating the linear relationship. \ Left inset: Transmission spectrum
of the microcavity with lorenzian fit to infer $\protect\omega _{m}$ and $%
Q_{m}$.}
\end{figure}

\begin{figure}[tbp]
\begin{center}
\includegraphics[
width=3.8374in
]{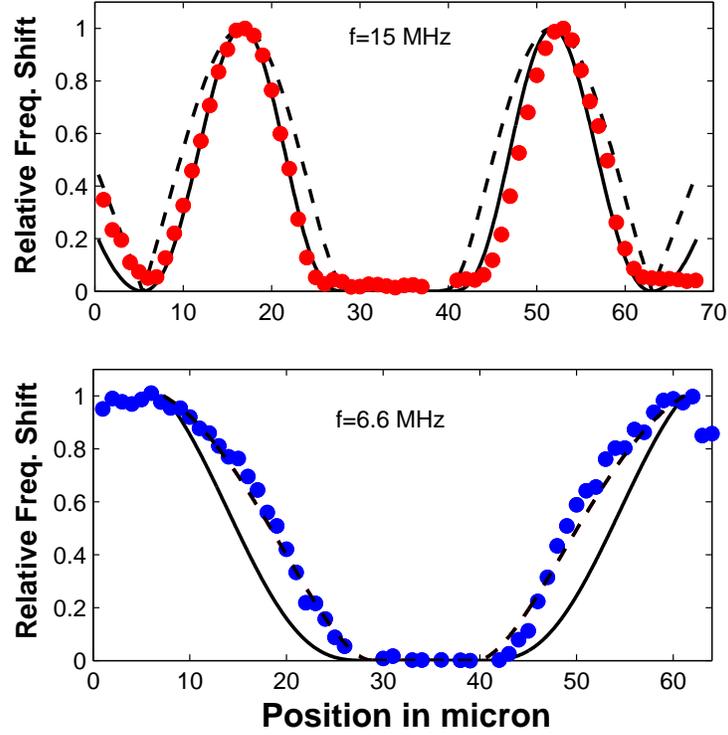}
\end{center}
\caption{ Scanning probe microscopy of the micro-mechanical resonances of
the $n=1$ and $n=2$ mode of the toroid microcavity shown in Fig 1b. Upper
graph: The normalized mechanical frequency shift for the $n=1$ mode as a
function of position. Lower graph: The normalized frequency shift for the $%
n=2$ mechanical mode as a function of scanned distance across the toroid.
Superimposed is the scaled amplitude (solid line) and the amplitude squared
(dotted line) of the mechanical oscillator modes obtained by finite element
simulation of the exact geometry parameters (as inferred by SEM).}
\end{figure}
\newpage 

\end{document}